\documentclass{egpubl}
\usepackage{egsr2026}
 
\SpecialIssuePaper         % uncomment for final version of Computer Graphics Forum, special issue
\CGFStandardLicense
\usepackage[T1]{fontenc}
\usepackage{dfadobe}  

\BibtexOrBiblatex
\electronicVersion
\PrintedOrElectronic
\ifpdf \usepackage[pdftex]{graphicx} \pdfcompresslevel=9
\else \usepackage[dvips]{graphicx} \fi

\usepackage{egweblnk} 
\usepackage{amssymb}
\usepackage{booktabs}
\usepackage{graphicx}
\usepackage{enumitem}
\usepackage{soul}
\usepackage{dsfont}
\usepackage{gensymb}
\usepackage{microtype}
\usepackage{pifont}
\usepackage{threeparttable}
\usepackage{collect}
\usepackage{xspace}
\usepackage[normalem]{ulem}
\usepackage{xfrac}
\usepackage{multirow}
\usepackage{wrapfig}
\usepackage{bbm}
\usepackage{makecell}
\usepackage[most]{tcolorbox}

\newcommand{\rev}[1]{\textcolor{black}{#1}}
\newcommand{\revv}[1]{\textcolor{black}{#1}}

\newcommand{\ie}{i.e.,\ }

\newcommand{\refFig}[1]{Fig.~\ref{fig:#1}}
\newcommand{\refTab}[1]{Tab.~\ref{tab:#1}}
\newcommand{\refSec}[1]{Sec.~\ref{sec:#1}}
\newcommand{\refEq}[1]{Eq.~\ref{eq:#1}}

\definecolor{lightgray230}{RGB}{230,230,230}
\newcommand{\coloredequation}[3]{
    \begin{tcolorbox}[
    colback=#1,
    colframe=#2,
    boxrule=0pt,
    arc=2pt,
    left=2pt,right=2pt,top=2pt,bottom=2pt
    ]
    \begin{equation}
    #3
    \end{equation}
    \end{tcolorbox}
}

\definecollection{mymaths}

\newcommand{\mymath}[2]{
    \newcommand{#1}{\TextOrMath{$#2$\xspace}{#2}}
    \begin{collect}{mymaths}{}{}{}{}
    #1
    \end{collect}
}

\mymath{\R}{\mathbb{R}}
\mymath{\dirspace}{\mathbb{S}}
\mymath{\pos}{\mathbf{x}}
\mymath{\dir}{\boldsymbol{\omega}}

\mymath{\voldomain}{B}
\mymath{\volboundary}{\partial\voldomain}
\mymath{\volnormal}{\mathbf{n}_\voldomain}
\mymath{\inflowboundary}{\Gamma}

\mymath{\radiance}{L}
\mymath{\radianceemission}{L_\mathrm{e}}
\mymath{\extinctioncoeff}{\sigma_\mathrm{t}}
\mymath{\absorptioncoeff}{\sigma_\mathrm{a}}
\mymath{\scatteringcoeff}{\sigma_\mathrm{s}}
\mymath{\phasefct}{f_\textrm{p}}

\mymath{\observation}{I}
\mymath{\imgindex}{k}
\mymath{\sceneindex}{l}
\mymath{\lightfieldnetparams}{\theta}
\mymath{\lightfieldnet}{\radiance_\lightfieldnetparams}
\mymath{\propertynetparams}{\phi}
\mymath{\propertynet}{M_\propertynetparams}
\mymath{\hgparam}{g}
\mymath{\loss}{\mathcal{L}}
\mymath{\lossweight}{\lambda}
\mymath{\inflowradiance}{\radiance_{\mathrm{in}}}

\mymath{\campos}{\mathbf{o}}
\mymath{\pix}{\mathbf{q}}

\mymath{\latent}{\mathbf{z}}
\mymath{\latentdim}{{d_{\mathbf{z}}}}

\mymath{\nerfnetparams}{\xi}
\mymath{\nerfnet}{F_\nerfnetparams}

\title{Volumetric Inverse Rendering via Neural Radiative Transfer}

\author[Nsampi et al.]
{
Ntumba Elie Nsampi$^1$, 
Adarsh Djeacoumar$^1$,
Hans-Peter Seidel$^1$,	
Tobias Ritschel$^2$	and
Thomas Leimkühler$^1$
\\
$^1$ Max-Planck-Institut für Informatik\qquad
$^2$ University College London
}

\begin{document}

\teaser{
 \includegraphics[width=0.99\linewidth]{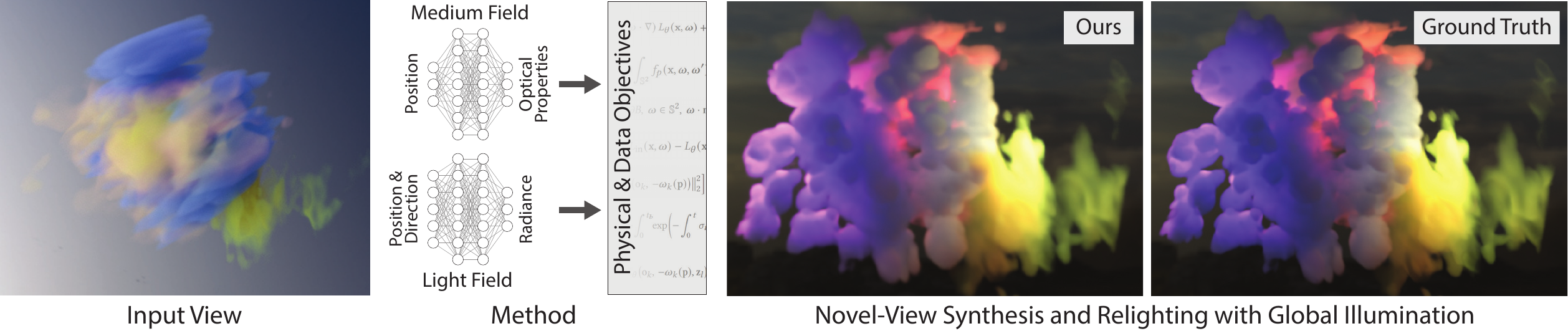}
 \centering
 \caption{We propose a formulation for volumetric inverse rendering under global illumination without explicit global-illumination rendering.
  By framing the problem as constrained optimization over neural fields, the approach enables physically consistent reconstruction, novel-view synthesis, and relighting. Here, we demonstrate relighting of the reconstructed volume \rev{with anisotropic optical properties} using a combination of environment illumination and three colored local light sources.}
\label{fig:teaser}
}

\maketitle

\begin{abstract}
Volumetric inverse rendering seeks to recover the optical properties of participating media from images.
Existing approaches either rely on differentiable stochastic light transport simulation, which require substantial algorithmic effort, or use simplified models that fail to capture global illumination.
We propose a formulation that reconciles physically complete light transport with general-purpose neural optimization.
The optical properties of the medium and the full light field are represented as neural fields and estimated through a joint optimization process.
Global illumination is enforced via a residual objective derived from the Radiative Transfer Equation in local differential form, complemented by a volume rendering term along primary viewing rays to mitigate \rev{low-frequency} bias.
We demonstrate reconstruction of spatially varying, color-resolved scattering, absorption, and phase function parameters from multi-view images.
Beyond reconstruction, the same framework supports learning generative models of participating media with physical optical properties under global illumination.

%  ACM CCS 2012
%(see https://www.acm.org/publications/class-2012)
%The tool at \url{http://dl.acm.org/ccs.cfm} can be used to generate
% CCS codes.
%Example:


\end{abstract}

%################################################################
%################################################################

\section{Introduction}

Light transport in volumetric scenes is complex.
As light propagates through participating media, it is continuously scattered and absorbed, often undergoing multiple scattering events.
These interactions generate long, intertwined transport paths that span the volume and tightly entangle illumination with the spatially varying optical properties of the medium.
The radiance observed at any point is therefore not the result of a small number of local interactions, but the cumulative outcome of a dense superposition of indirect light contributions.
Computational models of this process typically rely on radiative transfer simulations that use stochastic sampling of light paths~\cite{pharr2023physically}.

\begin{figure}
    \includegraphics[width=0.99\linewidth]{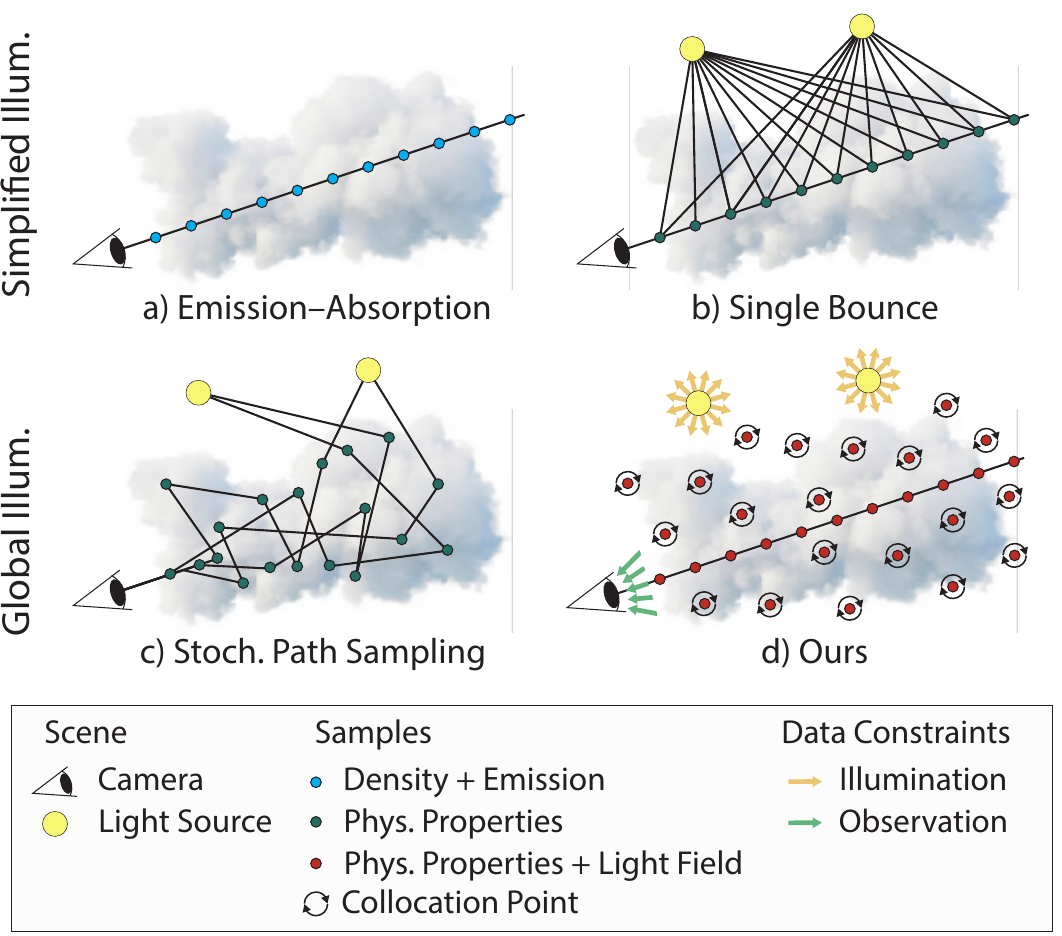}
    \vspace{-0.1cm}
    \caption{
        Different approaches to volumetric inverse rendering.
        \emph{(a)} The emission--absorption model bakes illumination into the volume, preventing recovery of physically meaningful optical properties.
        \emph{(b)} Restricting light transport to a single bounce enables limited relighting but fails to capture the complexity of global illumination.
        \emph{(c)} Differentiable stochastic path sampling accounts for global illumination during reconstruction, but requires substantial methodological and engineering effort.
        \emph{(d)} Our approach enables global-illumination-aware reconstruction with only minimal explicit light transport modeling by jointly optimizing the medium's physical properties and the scene's light field in a neural representation.
        Illumination and observations are incorporated via data constraints on the light field within a physics-informed formulation, while local light transport constraints are enforced at continuously sampled collocation points.
        To capture high-frequency detail, an integral volume rendering formulation is applied along primary viewing rays only.
    }
    \label{fig:positioning}
    \vspace{-0.5cm}
\end{figure}

Here, we study the corresponding inverse problem: given image observations of a scene containing participating media, we aim to reconstruct its spatially varying optical properties.
This challenging task is commonly formulated as an optimization problem, in which volumetric properties are estimated by minimizing a reconstruction objective using gradient-based methods.
Existing approaches can be broadly divided into two categories.
First, stochastic light path sampling is made differentiable~\cite{nimier2019mitsuba,li2018differentiable}, enabling the optimization to account for full radiative transfer~(\refFig{positioning}c).
However, these methods demand substantial methodological and engineering effort to achieve acceptable efficiency~\cite{nimier2020radiative,vicini2021path,nicolet2023recursive} and to control bias in the estimated gradients~\cite{nimier2022unbiased}.
In a second line of work, simplified models of light transport are employed, often in combination with neural scene representations~\cite{mildenhall2020nerf,srinivasan2021nerv,zheng2021neural,zhang2023nemf}~(\refFig{positioning}a,\,b).
While these approaches benefit from general-purpose, easy-to-use machine-learning frameworks, they typically fail to capture accurate optical properties under global illumination.
To date, these two lines of work have remained largely disjoint.

In this work, we explore a way to \emph{reconcile the generality of physically complete light transport formulations with \rev{the flexibility and practicality of modern} machine learning}.
Our approach shifts complexity from transport-specific algorithm design to learned representations and optimization, thereby decoupling physical modeling from specialized numerical treatment.

Specifically, we represent both the optical properties of the participating media and the full light field of the scene using neural fields~\cite{xie2022neural}, and formulate their joint estimation as a continuous optimization problem~(\refFig{teaser}).
Crucially, this optimization involves only minimal explicit simulation of light transport~(\refFig{positioning}d).
Global illumination is incorporated through a residual objective derived from the Radiative Transfer Equation (RTE), following a physics-informed formulation based on local differential operators~\cite{raissi2019physics,hadadan2021neural,hadadan2023inverse}.
Although the supervision signals induced by the RTE are local, their propagation through optimization implicitly resolves the dense superposition of light paths, leading the joint estimate of the light field and the medium's optical properties to converge to a globally consistent light transport equilibrium without explicit transport modeling.
The input observations and illumination conditions are incorporated through corresponding data constraints.
In addition, we include a data term based on the Volume Rendering Equation (VRE), which integrates radiance along primary viewing rays only.
This term mitigates the well-known \rev{low-frequency} bias of physics-informed neural networks~\cite{wang2022when,rahaman2019spectral} by coupling the reconstructed light field to view-ray integrals \rev{(we avoid the term "spectral bias", commonly used in the machine learning literature, to prevent confusion with the wavelength dependence of light)}.
Importantly, explicit higher-order light transport simulation is avoided, since the RTE-based formulation accounts for global illumination.
This substantially decouples inverse rendering research from renderer-specific algorithm design.

Using our formulation, we demonstrate the feasibility of reconstructing spatially varying, color-resolved scattering, absorption, and phase function parameters from multi-view observations under known illumination conditions, with competitive performance.
Moreover, because the formulation relies on general-purpose neural optimization rather than transport-specific algorithmic machinery, it naturally extends
beyond reconstruction to generative modeling of participating media with physically meaningful optical properties under full global illumination, learned directly from image observations. 
% \revv{Our formulation readily supports latent conditioning, enabling the model to represent distributions over scenes rather than a single fixed reconstruction. By jointly modeling medium properties and light fields, we exploit light transport structure across scenes instead of requiring independent transport simulation for each sample.}
% Moreover, because the formulation relies on general-purpose neural optimization rather than transport-specific algorithmic machinery, it naturally extends beyond reconstruction to generative modeling of participating media with physically meaningful optical properties under full global illumination, learned directly from image observations.
In summary, our contributions are:
\begin{itemize}
    \item A novel formulation for volumetric inverse rendering that accounts for full global illumination with only minimal reliance on transport-specific simulation.
    \item A joint estimation framework for a scene’s optical properties and full light field, both represented as neural fields and constrained by first principles of global light transport.
    \item Experimental demonstrations of optical property reconstruction and the application of the formulation to generative modeling of physically meaningful participating media.
\end{itemize}

All source code and trained models are available at : \href{https://neural-radiative-transfer.mpi-inf.mpg.de/}{https://neural-radiative-transfer.mpi-inf.mpg.de/}

%################################################################
%################################################################

\section{Related Work}

Here, we review related work on \revv{inverse light transport~(\refSec{inv-rte})}, physics-informed learning~(\refSec{pinn}), and physical generative models~(\refSec{related-generative}).

% inverse rendering~(\refSec{vol-inverse-rendering})
%################################################################

\subsection{Inverse Light Transport}
\label{sec:inv-rte}

%\subsubsection{Traditional Transport-Based Inverse Radiative Transfer}
\subsubsection{\revv{Discretization-Based Inverse Radiative Transfer}}

% The unknown optical properties are inferred by iteratively solving the discretized forward problem, where the radiative transfer equation serves as the governing physics constraint.
Inverse radiative transfer problems have traditionally been solved by numerically discretizing the radiative transfer equation~\cite{mccormick1992inverse, adams2002fast}. Traditional methods typically represent the light field and the optical properties using linear basis expansions defined on a computational mesh, resulting in a finite-dimensional approximation of the governing equation~\cite{bangerth2008adaptive}. 
The optical properties are then inferred by iteratively solving the resulting system.
These approaches have been successfully applied in fields such as atmospheric remote sensing~\cite{levis2015airborne, levis2017multiple},  medical imaging~\cite{abdoulaev2005optical} and constitute the foundation of many established radiative transfer frameworks~\cite{stamnes2000disort}.
While traditional methods rely on discretization, our proposed method is based on a physics-informed learning framework, where we represent both the light field and the optical properties as neural networks.
The governing physics and observational constraints are then enforced through the optimization objective.

%################################################################

% \subsection{Inverse Rendering and Differentiable Light Transport}
% \label{sec:vol-inverse-rendering}

\subsubsection{Differentiable Monte Carlo Light Transport}

%Early work on inverse volume rendering optimized physical optical properties of homogeneous participating media using physically based Monte Carlo light transport, often relying on hand-crafted priors~\cite{gkioulekas2013inverse}.

\rev{Inverse volume rendering is traditionally addressed by differentiating stochastic light transport simulations, typically based on path tracing, to optimize the optical properties of participating media~(\refFig{positioning}c).}
Gradients of image measurements with respect to scene parameters are estimated by differentiating Monte Carlo estimators of the rendering equation, enabling optimization under physically complete light transport~\cite{nimier2019mitsuba,zhang2019differential,zhang2021path}.
This framework has been applied to volumetric and translucent materials, allowing inverse estimation of optical properties in participating media~\cite{gkioulekas2013inverse,gkioulekas2016evaluation,deng2022reconstructing,deng2024reconstructing,weier2025practical}.
While conceptually general, differentiable Monte Carlo methods are computationally expensive and often limited by high gradient variance, motivating extensive work on improving efficiency and stability~\cite{nimier2020radiative,vicini2021path,nicolet2023recursive}.
%Moreover, despite advances in handling sampling discontinuities, particularly in surface-based settings~\cite{loubet2019reparameterizing,li2018differentiable}, controlling bias of estimated gradients remains an active area of research~\cite{nimier2022unbiased,Worchel2025RadiativeBackpropagation,zeltner2021monte}.
\rev{Moreover, multiple scattering requires differentiating an expectation over a stochastic branching process. 
Obtaining gradients which are useful in optimization requires substantial methodological and engineering effort and remains an active area of research~\cite{nimier2022unbiased,Worchel2025RadiativeBackpropagation,zeltner2021monte}.}
Our approach sidesteps these challenges by avoiding stochastic \rev{path sampling} altogether, while still accounting for physically complete light transport.

%################################################################

%\subsubsection{Learning-Based Inverse Rendering}

%Learning-based approaches to inverse rendering commonly adopt simplified light transport models.

\subsubsection{\rev{Inverse Rendering with Learned Representations}}
\rev{Inverse rendering approaches based on learned representations} commonly adopt simplified light transport models.
A prominent example is the emission--absorption formulation~(\refFig{positioning}a), which represents volumes as purely emissive and neglects explicit illumination and scattering~\cite{mildenhall2020nerf,kerbl3Dgaussians,kajiya1984ray}, \rev{making it amenable to automatic differentiation without special treatment.}
Single-scattering approaches constitute the next level of modeling complexity by accounting for direct illumination and a single light bounce~(\refFig{positioning}b).
A common paradigm in this setting is to infer surface-based material properties, such as albedo and normals~\cite{zhang2021nerfactor,zhang2021physg,li2024tensosdf,bi2020neural,liang2024gs,dihlmann2024subsurface}.
%removed gao2024relightable to fit in one line
Accuracy can be further improved by explicitly modeling incident illumination using specialized components~\cite{yao2022neilf,boss2021neural,srinivasan2021nerv,rng2025} and by introducing physics-inspired regularization terms that encourage consistency with simplified reflectance models~\cite{wu2025pbr}.
Recent work has also explored learning-based solutions for volumetric participating media with anisotropic scattering~\cite{zhang2023nemf}, but remains limited to restricted transport models.
Multi-bounce illumination in inverse rendering has been considered more rarely.
In this setting, surface-based representations can leverage precomputed radiance transfer~\cite{lyu2022nrtf}, caching techniques~\cite{shi2025gir}, or explicit global-illumination decomposition strategies~\cite{chen2025gigs,jin2023tensoir}.
For volumetric scenes, however, existing approaches typically rely on strong assumptions, such as homogeneous media~\cite{che2020towards}, or on simplified treatments of higher-order light transport~\cite{zheng2021neural}.
In contrast, our approach does not rely on such restrictive assumptions and directly targets physically complete volumetric light transport.

%################################################################

\subsection{Physics-Informed Learning for Light Transport and Scene Reconstruction}
\label{sec:pinn}

Physics-informed neural networks (PINNs) integrate physical laws, typically expressed as (integro-)differential equations, into neural network training by penalizing violations of these equations in the loss function~\cite{raissi2019physics}.
Incorporating the governing equations of volumetric light transport into such frameworks has been explored in both forward and inverse settings~\cite{mishra2021physics,zucker2025physics,riganti2023auxiliary}.
\rev{These works focus on ultra low-frequency optical properties and light fields, and often are further restricted to low-dimensional (1D or 2D) settings, whereas our approach operates in full 3D and aims to reconstruct spatially complex, high-frequency volumetric structure, as required in inverse graphics.}
This is achieved using an explicit integration scheme along primary rays, which helps alleviate the \rev{low-frequency} bias of neural networks in general~\cite{rahaman2019spectral} and of PINN-based formulations in particular~\cite{wang2022when}.

Physics-informed learning has also been explored for light transport in surface-based scenes, in both forward~\cite{hadadan2021neural} and inverse~\cite{hadadan2023inverse} settings.
In these surface-based formulations, non-local transport through free space is naturally handled via stochastic ray sampling, whereas our volumetric formulation enforces light transport locally throughout the volume \revv{by imposing the RTE as a pointwise constraint.}
\revv{Optimizing hard geometry, \ie surfaces with sharp discontinuities in PINN-based inverse rendering remains challenging and has not yet been demonstrated, whereas our method naturally supports the optimization of soft geometry (e.g., continuous volumetric media such as clouds or fog).}

\rev{Another line of work addresses the neural reconstruction of dynamic volumes by incorporating physical soft constraints derived from fluid dynamics~\cite{chu2022physics,yu2024inferring}.
While these approaches model the transport of matter in participating media, our work instead focuses on the transport of light.}

%################################################################

\subsection{Generative Models of Optical Properties}
\label{sec:related-generative}

Recent work has explored generative modeling of scene appearance and optical properties, typically under simplified image formation models.
A first line of work focuses on surface-based representations and learns generative models of material appearance under simple shading assumptions, often limited to direct illumination~\cite{pan2021shading,violante2024lighting_3D_cars,jiang2023nerffacelighting,chen2023fantasia3d,zhang2024dreammat}.
These methods do not model volumetric light transport or participating media.
A second line of work considers volumetric scene representations and learns generative models under emission--absorption rendering~\cite{henzler2019escaping,diolatzis2023mesogan,chan2022efficient,muller2023diffrf,chen2023single}.
While some approaches target volumetric phenomena such as participating media, the representations remain appearance-driven and do not explicitly model physical optical parameters or enforce global illumination.
To our knowledge, no prior work enables generative modeling of physical volumetric optical properties, such as absorption, scattering, and phase functions, under full global illumination.

%Tangentially related work repurposes large foundation models to extract coarse optical properties~\cite{zeng2024rgb}, decompose material and illumination~\cite{lyu2023dpi}, or modify lighting~\cite{zeng2024dilightnet,erel2025practilight}.
%These approaches primarily operate on surface-based representations and image-space appearance, and do not constitute generative modeling of volumetric optical properties.

%################################################################
%################################################################

\section{Radiative Transfer Preliminaries}
\label{sec:background}

The full volumetric light transport in scenes with participating media is governed by the Radiative Transfer Equation (RTE)~\cite{chandrasekhar1960radiative,pharr2023physically}, an integro-differential equation that describes how radiance evolves through the medium under absorption, emission, and scattering.
We denote by $\radiance(\pos,\dir) \in \R^3$ the radiance at spatial location $\pos \in \R^3$ traveling in direction $\dir \in \dirspace^2$, which characterizes the spatio-angular distribution of light throughout the volume.
We refer to this function as \emph{light field}, avoiding the term \emph{radiance field} to prevent confusion with emission--absorption-based scene representations~\cite{mildenhall2020nerf}, in which ``radiance'' denotes a local appearance term that is conceptually closer to \radianceemission than to the steady-state radiance \radiance satisfying the RTE.
We model the spectral dependence of the light~field and of all other quantities using three (RGB) channels.

Under the assumption of steady state, the RTE takes the form
\begin{equation}
\begin{aligned}
\overbrace{
    (\dir \cdot \nabla)\,\radiance(\pos, \dir)
}^{\textrm{Transport}}
&=
-
\overbrace{
    \extinctioncoeff(\pos)\,\radiance(\pos, \dir)
}^{\textrm{Extinction}}
+
\overbrace{
    \absorptioncoeff(\pos)\,\radianceemission(\pos, \dir)
}^{\textrm{Emission}}
\\
&\quad
+
\underbrace{
    \scatteringcoeff(\pos)
    \int_{\dirspace^2}
    \phasefct(\pos, \dir, \dir')\,\radiance(\pos, \dir')\,\mathrm{d}\dir'}
_{\textrm{In-scattering}}\!\!.
\end{aligned}
\label{eq:rte}
\end{equation}
The medium is parameterized by spatially-varying RGB absorption and scattering coefficients 
$\absorptioncoeff(\pos), \scatteringcoeff(\pos) \in \R^3$, 
with the extinction coefficient given by 
$\extinctioncoeff(\pos)=\absorptioncoeff(\pos)+\scatteringcoeff(\pos)$.
Volumetric emission is described by the emitted radiance 
$\radianceemission(\pos,\dir) \in \R^3$.
Scattering is governed by the spatially varying phase function 
$\phasefct(\pos,\dir,\dir') \in \R_{\geq 0}$, 
which specifies the normalized angular redistribution of light scattered from incident direction $\dir'$ into outgoing direction \dir.
The left-hand side of \refEq{rte} describes infinitesimal directional transport of radiance through the medium, while the right-hand side accounts for local loss due to extinction and local gain due to emission and in-scattering.
Although the scattering term integrates radiance over directions, all interactions in the RTE are local in space.
Through this angular coupling induced by scattering, radiance at any location and direction arises from a dense superposition of indirect light transport contributions spanning the entire volume, corresponding to full global illumination.

% An explicit equation for $\radiance(\pos,\dir)$ can be obtained by integrating both sides of \refEq{rte}, giving the Volume Rendering Equation (VRE)~\cite{kajiya1984ray,pharr2023physically}:
An explicit equation for \radiance(\pos,\dir) can be obtained by integrating both sides of \refEq{rte}. \revv{This integral form has a long history in radiative transfer physics~\cite{chandrasekhar1960radiative}, and is known in computer graphics as the Volume Rendering Equation (VRE)~\cite{kajiya1984ray,pharr2023physically}:} 
\begin{equation}
\begin{aligned}
\radiance(\pos,\dir)
&=
\int_{0}^{t_b}
\overbrace{
\exp\!\left(
    - \int_{0}^{t} \extinctioncoeff(\pos_s)\,\mathrm{d}s
\right)
}^{\textrm{Transmittance}}
\Bigg[
\overbrace{
    \absorptioncoeff(\pos_t)\,\radianceemission(\pos_t,\dir)
    \vphantom{
        \exp\!\left(
            - \int_{0}^{t} \extinctioncoeff(\pos_s)\,\mathrm{d}s
        \right)
    }
}^{\textrm{Emission}}
\\
&\qquad\quad
+
\underbrace{
    \scatteringcoeff(\pos_t)
    \int_{\dirspace^2}
    \phasefct(\pos_t,\dir,\dir')\,
    \radiance(\pos_t,\dir')\,\mathrm{d}\dir'
}_{\textrm{In-scattering}}
\Bigg]
\,\mathrm{d}t.
\end{aligned}
\label{eq:vre}
\end{equation}
Here, points along the ray $\pos + \tau\,\dir$ are denoted by $\pos_\tau$ for $\tau \ge 0$, and $t_b$ is the distance to the volume boundary.
The volume rendering equation expresses radiance along a ray as an integral over local source terms, with the local extinction coefficient accumulated into a transmittance factor that encodes cumulative attenuation.

Both the differential RTE and the integral VRE are inherently recursive due to scattering-induced self-coupling of radiance; however, this recursion is expressed implicitly in the RTE through spatially local gain--loss constraints, whereas in the VRE it appears explicitly through nested, spatially nonlocal ray integrals.

%################################################################
%################################################################

\section{Method}

Input to our method is a set of posed RGB images $\{\observation_\imgindex\}$ providing multi-view observations of a scene containing \rev{heterogeneous} participating media.
We assume a known and fixed environment illumination.
Our goal is to reconstruct the optical properties of the participating media, namely spatially varying, color-resolved absorption and scattering coefficients $\absorptioncoeff(\pos)$ and $\scatteringcoeff(\pos)$, as well as spatially varying phase functions $\phasefct(\pos,\dir,\dir')$, under physically complete volumetric light transport.

While our formulation focuses on reconstructing the optical properties of a single scene, it naturally generalizes to a generative setting when observations from multiple scenes are available.
\rev{Such a generative model can be used to populate virtual scenes with a diverse set of assets, such as clouds. 
It could also be employed in an inverse setting, where a latent code is optimized to explain an observation (a photo of a participating medium), thereby providing a strong prior on the distribution of plausible 3D volumes.}
%, in which case we learn a scene-level distribution over spatially varying optical properties, where each sample corresponds to a coherent volumetric scene.
In both settings -- reconstruction and generation -- the recovered physical scene representations enable physically meaningful analysis, novel-view synthesis, relighting under novel illumination, and other operations that rely on interpretable optical properties.

Physically accurate light transport in participating media involves complex, highly coupled interactions~(\refSec{background}) that are typically addressed using specialized simulation techniques.
Here, we instead explore how scene properties can be estimated under these interactions through general-purpose neural optimization, with only minimal reliance on explicit transport simulation.

We first introduce our general framework for volumetric inverse rendering (\refSec{rep+opt}).
We then describe how this approach can be extended to the generative setting (\refSec{generative}), before providing implementation details (\refSec{implementation}).

%################################################################

\begin{figure}
    \includegraphics[width=0.99\linewidth]{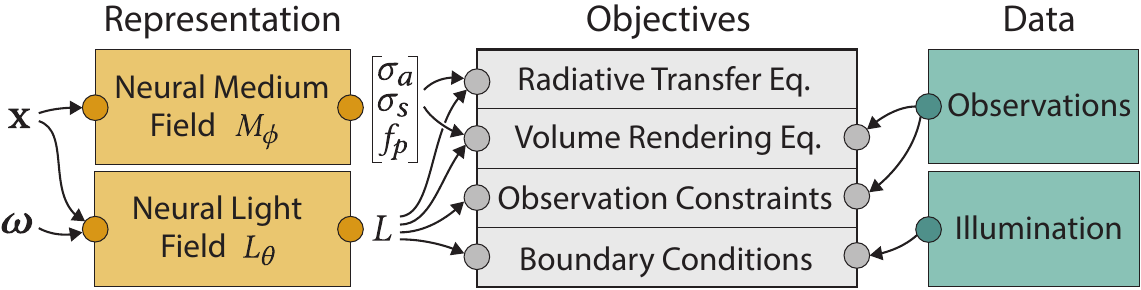}
    \vspace{-0.1cm}
    \caption{
       Overview of our approach.
        Two neural fields encode the optical properties of the participating medium and the scene's light field (orange).
        Disentanglement is guided by multiple optimization objectives (grey) that incorporate the available data (green).
        %While our ultimate goal is the recovery of the optical properties encoded in the medium network \propertynet, the scene's light field is introduced as an auxiliary representation and optimized jointly to facilitate their disentanglement under global illumination.
    }
    \vspace{-0.3cm}
    \label{fig:overview}
\end{figure}

\subsection{Neural Representation and Optimization}
\label{sec:rep+opt}

Central to our approach is the representation of the light field and the optical properties of the participating medium as neural fields~(\refFig{overview},~left).
Specifically, we employ a neural field
$\lightfieldnet : (\pos,\dir) \rightarrow \R^3$,
which encodes the spatio-angular light field of the entire scene~\cite{sitzmann2021light,keller2019integral},
and a neural field
$
\propertynet :
\pos \rightarrow
\left( \absorptioncoeff, \scatteringcoeff, \hgparam \right)
$,
which encodes the participating medium through spatially varying absorption and scattering coefficients and a scalar parameter \hgparam that defines a spatially varying Henyey--Greenstein~\cite{henyey1941diffuse} phase function \phasefct.
The subscripts \lightfieldnetparams and \propertynetparams denote the optimizable parameters of the respective neural fields.
To keep the notation concise, we omit explicit parameter subscripts for the medium properties in the remainder of this section and implicitly treat \absorptioncoeff, \scatteringcoeff, and \phasefct as outputs of \propertynet.

Disentanglement of the light field and medium properties, as well as the incorporation of data, is achieved through physics-informed optimization objectives combined with data constraints~(\refFig{overview}, right), rather than through specialized architectural mechanisms for separating illumination and material properties.
Accordingly, the remainder of this section details the individual constraints that form the core of our approach.
While our ultimate objective is the recovery of the output of \propertynet, most optimization objectives are imposed on \lightfieldnet to facilitate disentanglement of illumination and medium.

The central objective enforcing physically consistent global light transport is the formulation of \refEq{rte} as a residual:
\coloredequation{lightgray230}{lightgray230}{
\scalebox{1}{$
\begin{aligned}
    \loss_{\mathrm{RTE}}(\lightfieldnetparams, \propertynetparams)
    =
    &\mathbb{E}_{\pos, \dir}
    \Bigg[
    \Big\|
        (\dir \cdot \nabla)\,\lightfieldnet(\pos, \dir)
        + \extinctioncoeff(\pos)\,\lightfieldnet(\pos, \dir)
        %\!\!\!\!\!
        %- \absorptioncoeff(\pos)\,\radianceemission(\pos, \dir)
        \\&
        - \scatteringcoeff(\pos)
        \!\int_{\dirspace^2}
        \phasefct(\pos, \dir, \dir')\,\lightfieldnet(\pos, \dir')\,\mathrm{d}\dir'
    \Big\|_2^2
    \Bigg].
\end{aligned}
$}
\label{eq:rte-residual}
}
\noindent
Here, we omit the emission term, since we consider only environment illumination, and $\absorptioncoeff(\pos)$, $\scatteringcoeff(\pos)$, and $\phasefct(\pos,\dir,\dir')$ 
are obtained from the neural medium field \propertynet.
Note that this formulation enforces the radiative transfer constraints at continuously sampled collocation points $(\pos,\dir)$ within the spatio-angular domain (\refFig{positioning}d) and is purely local in space.
The first term in the expectation is evaluated using automatic differentiation.
We approximate the in-scattering integral using Monte Carlo sampling over the angular domain, evaluated locally at each collocation point’s spatial location $\pos$, without tracing rays or explicitly tracking light through the volume.

%Near-field light sources are incorporated through the emission term $\radianceemission$ in \refEq{rte-residual}~(orange arrows in \refFig{positioning}d).
To ensure a unique solution of the radiative transfer equation, the light field is additionally constrained by inflow boundary conditions at the volume boundary.
Let $\voldomain \subset \mathbb{R}^3$ denote the spatial domain of the scene, with boundary \volboundary and outward unit normal $\volnormal(\pos)$.
The inflow boundary \inflowboundary consists of all boundary position--direction pairs for which radiance enters the domain:
\begin{equation}
    \inflowboundary
    =
    \left\{
    (\pos,\dir)\ \middle|\ 
    \pos \in \volboundary,\ 
    \dir \in \dirspace^2,\ 
    \dir \cdot \volnormal(\pos) < 0
    \right\}.
\label{eq:inflowboundary}
\end{equation}
Boundary conditions are enforced through
\coloredequation{lightgray230}{lightgray230}{
\loss_{\mathrm{BC}}(\lightfieldnetparams)
    =
    \mathbb{E}_{(\pos,\dir) \in \inflowboundary}
    \Big[
    \big\|
     \inflowradiance(\pos,\dir) - \lightfieldnet(\pos,\dir)
    \big\|_2^2
    \Big],
\label{eq:bc-loss}
}
\noindent
where \inflowradiance represents the incoming radiance at the boundary originating from environment illumination.

So far, we have only enforced data-agnostic physical and boundary constraints within the optimization.
To incorporate the available image data, we introduce additional constraints on the light field derived from the observations.
For an image $\observation_{\imgindex}$, each pixel $\pix \in \R^2$ corresponds to a viewing direction 
$\dir_{\imgindex}(\pix) \in \dirspace^2$ 
originating at the camera center 
$\campos_{\imgindex} \in \R^3$.
Under known camera calibration, the observed pixel value 
$\observation_{\imgindex}(\pix) \in \R^3$ 
specifies the radiance arriving at the camera location from the corresponding viewing direction, \ie traveling along direction $-\dir_{\imgindex}(\pix)$ (green arrows in \refFig{positioning}d).
We therefore impose this as a pointwise constraint through
\coloredequation{lightgray230}{lightgray230}{
    \loss_{\mathrm{obs}}(\lightfieldnetparams)
    =
    \mathbb{E}_{\imgindex,\pix}
    \Big[
    \big\|
    \observation_{\imgindex}(\pix)
    -
    \lightfieldnet\big(\campos_{\imgindex},\,-\dir_{\imgindex}(\pix)\big)
    \big\|_2^2
    \Big],
    \label{eq:obs-loss}
}
\noindent
where the expectation is taken over all observed views and pixels.
This term constrains the light field only at the camera locations and for the set of observed incoming directions.

While the three objectives introduced so far are theoretically sufficient to guide the joint optimization of the light field \lightfieldnet and the material field \propertynet, in practice we observe that the resulting solutions are overly smooth and lack high-frequency detail.
This aligns with previous observations that neural network–based representations exhibit a \rev{bias} toward low-frequency solutions even under direct supervision~\cite{rahaman2019spectral}, and that related optimization pathologies have been reported for physics-informed formulations relying on differential constraints~\cite{wang2022when}.
We alleviate this effect by introducing an additional objective based on the VRE (\refEq{vre}), which explicitly integrates radiance along primary viewing rays (ray in \refFig{positioning}d) and thereby introduces nonlocal constraints that counteract the oversmoothing induced by purely local supervision:
\coloredequation{lightgray230}{lightgray230}{
    \scalebox{0.98}{$
    \begin{aligned}
    \loss_{\mathrm{VRE}}(\lightfieldnetparams,&\propertynetparams)= 
    \mathbb{E}_{\imgindex,\pix}
    \Bigg[
    \Bigg\|
    \observation_{\imgindex}(\pix)
    - 
    \!\int_{0}^{t_b}
    \!\!\exp\!\left(
        - \int_{0}^{t}
        \extinctioncoeff(\pos_s)\,\mathrm{d}s
    \right)
    \\
        %\absorptioncoeff(\pos_t)\,
        %\radianceemission\!\big(\pos_t, -\dir_{\imgindex}(\pix)\big)
    &
    %\Bigg(
        \scatteringcoeff(\pos_t)
        \!\int_{\dirspace^2}
        \phasefct\!\big(\pos_t,-\dir_{\imgindex}(\pix),\dir'\big)\,
        \lightfieldnet(\pos_t,\dir')\,\mathrm{d}\dir'
    %\Bigg)
    \mathrm{d}t
    \Bigg\|_2^2
    \Bigg],
    \end{aligned}
    $}
\label{eq:vre-loss}
}
\noindent
where the locations $\pos_\tau$ lie on the primary viewing ray
$\campos_{\imgindex} + \tau\,\dir_{\imgindex}(\pix)$, while radiance is evaluated in the incoming direction $-\dir_{\imgindex}(\pix)$.

Crucially, the in-scattering radiance in \refEq{vre-loss} is obtained by querying the neural light field \lightfieldnet, which is jointly optimized under all objectives, with global light transport consistency enforced by the RTE residual \refEq{rte-residual}.
As a result, although \refEq{vre-loss} aggregates radiance only along primary viewing rays, the radiance values integrated during volume rendering are not restricted to single-scattering or direct illumination, but instead evolve to encode the cumulative effect of arbitrarily many scattering interactions throughout the volume.
Intuitively, the RTE governs the accumulation of indirect illumination throughout the volume, while the VRE term provides complementary non-local data coupling that anchors the directly observed component along each viewing direction.

In practice, the integrals in \refEq{vre-loss} are approximated using stochastic sampling.
The line integral along each viewing ray is evaluated using jittered stratified sampling, following standard practice~\cite{pauly2000metropolis,mildenhall2020nerf}.
The directional in-scattering integral appearing in \refEq{vre-loss} is handled identically to the corresponding term in the RTE residual \refEq{rte-residual}, using Monte Carlo sampling over the angular domain and evaluated locally at each spatial location.

Our complete optimization objective is given by
\begin{equation}
\begin{aligned}
    \loss(\lightfieldnetparams,\propertynetparams)
    &=
    \loss_{\textrm{RTE}}(\lightfieldnetparams,\propertynetparams)
    + \lossweight_{\textrm{BC}} \loss_{\textrm{BC}}(\lightfieldnetparams)
    \\
    &\,+ \lossweight_{\textrm{obs}} \loss_{\textrm{obs}}(\lightfieldnetparams)
    + \lossweight_{\textrm{VRE}} \loss_{\textrm{VRE}}(\lightfieldnetparams,\propertynetparams),
\end{aligned}
\label{eq:full-loss}
\end{equation}
where the scalar weights $\lossweight_{\textrm{BC}}$, $\lossweight_{\textrm{obs}}$, and $\lossweight_{\textrm{VRE}}$ balance the relative contributions of the individual objectives.

%################################################################

\subsection{Extension to Generative Modeling}
\label{sec:generative}

Beyond inverse rendering of individual scenes, our approach naturally extends to learning a generative model of volumetric scene properties from a distribution of observations across \emph{multiple} scenes.
Specifically, we assume access to an extended set of posed observations
$\{\observation_{\sceneindex,\imgindex}\}$,
where \sceneindex indexes scenes and \imgindex indexes images within each scene, as before.
In our proof-of-concept application, we assume the same illumination conditions across scenes.

We employ an auto-decoder framework~\cite{park2019deepsdf}, where each scene is associated with an optimizable latent code
$\latent_{\sceneindex} \in \R^{\latentdim}$.
This latent code is provided as an additional input to both \lightfieldnet and \propertynet, conditioning them on the corresponding scene.
Training proceeds as in the scene-specific case described in \refSec{rep+opt}, with only two changes.
First, the scene-specific latent codes 
$\{\latent_{\sceneindex}\}$ are jointly optimized with the neural field parameters \lightfieldnetparams and \propertynetparams, \rev{and are additionally regularized with an $L_2$ penalty~\cite{park2019deepsdf}.}
Second, the observation-related objectives $\loss_\mathrm{obs}$ (\refEq{obs-loss}) and $\loss_\mathrm{VRE}$ (\refEq{vre-loss}) are extended to account for the distribution over scenes.
\refEq{obs-loss} now reads
\begin{equation}
    \loss_{\mathrm{obs}}(\lightfieldnetparams, \{\latent_{\sceneindex}\})
    =
    \mathbb{E}_{\sceneindex, \imgindex,\pix}
    \Big[
    \big\|
    \observation_{\sceneindex, \imgindex}(\pix)
    -
    \lightfieldnet\big(\campos_{\imgindex},\,-\dir_{\imgindex}(\pix), \latent_{\sceneindex}\big)
    \big\|_2^2
    \Big],
    \label{eq:obs-loss-generative}
\end{equation}
and \refEq{vre-loss} is modified analogously.
The resulting model defines a generative distribution over volumetric scenes, jointly capturing light fields and material property fields, which can be sampled by randomly drawing a latent code \latent and evaluating the corresponding neural fields to obtain a complete scene.

It is worth noting that alternative inverse rendering formulations, including differentiable stochastic light transport, could in principle also be extended to learn distributions over medium properties, provided that the underlying scene representation supports such variability.
However, doing so would require an independent light transport simulation for each sampled scene instance and for each training iteration, incurring the full cost of global illumination evaluation repeatedly during optimization.
In contrast, our formulation jointly models a distribution over both medium properties and light fields, allowing the optimization to exploit shared structure across scenes and to amortize the resolution of global light transport through the learned light field representation.
While the light field remains an auxiliary quantity, this joint modeling enables efficient learning of scene-level variability under physically complete volumetric illumination.

%################################################################

\subsection{Implementation Details}
\label{sec:implementation}

Our implementation relies exclusively on general-purpose machine-learning components provided by PyTorch~\cite{paszke2017automatic}, without requiring any custom operations or rendering-specific kernels.
We minimize \refEq{full-loss} using the Adam~\cite{kingma2014adam} optimizer with default settings, and 
$\lossweight_{\textrm{BC}} = \lossweight_{\textrm{obs}}=1$, and
$\lossweight_{\textrm{VRE}}=70$. 

As observed in prior work on both differentiable sampling-based and learning-based inverse rendering~\cite{nimier2022unbiased,zhang2021nerfactor}, scene initialization plays a crucial role in obtaining high-quality solutions.
Accordingly, following common practice, we adopt a two-stage optimization procedure.
In the first stage, we train an emission--absorption-based reconstruction model \nerfnet~\cite{mildenhall2020nerf} (\refFig{positioning}a) using a neural hash grid architecture~\cite{muller2022instant}.
The color output of \nerfnet is discarded, as it represents a non-physical emission term.
In the second stage, we jointly optimize the neural medium field \propertynet and the light field \lightfieldnet, both parameterized by SIREN networks~\cite{sitzmann2020implicit}.
The medium field \propertynet multiplicatively modulates the density output of \nerfnet, which provides a strong high-frequency initialization for the volumetric structure.
\rev{We enforce positivity of \scatteringcoeff and \absorptioncoeff via an exponential mapping, and bound \hgparam using a hyperbolic tangent.
Experiments using an albedo--extinction parameterization yielded slightly inferior results.}
In the generative setting, latent codes \latent are incorporated differently, with \nerfnet using multiplicative modulation of its hash grid features via an MLP-based encoding of \latent prior to decoding, and \propertynet and \lightfieldnet concatenating the latent code with the positional and directional inputs.
\rev{The described architectural design choices yielded the best results in a reasonably comprehensive pilot search.}

In all experiments, we use 10k collocation points to evaluate \refEq{rte-residual}, 32 directional samples to approximate the in-scattering integrals in \refEq{rte-residual} and \refEq{vre-loss}, 10k boundary samples for \refEq{bc-loss}, and 100 samples per viewing ray for evaluating \refEq{vre-loss}.
All samples are drawn randomly from the corresponding continuous domains at each optimization iteration.
To avoid optimization gradient bias arising from stochastic sampling of in-scattering directions, we draw independent samples for the forward and backward passes~\cite{azinovic2019inverse}.
\rev{The runtime of our method is almost entirely attributable to evaluating and optimizing the four objective terms $\loss_{\mathrm{RTE}}$ (38\%), $\loss_{\mathrm{BC}}$ (7\%), $\loss_{\mathrm{obs}}$ (2\%), and $\loss_{\mathrm{VRE}}$ (53\%).}

%################################################################
%################################################################

\begin{table*}
\begin{center} 
\caption
{
    Quantitative evaluation of different methods (rows) across multiple evaluation protocols (columns) for scenes with \emph{isotropic} phase functions. We report the accuracy of the recovered optical properties, the quality of novel-view synthesis under novel illumination, \rev{optimization time when using recommended hyperparameters} on a single A100 GPU, and the generative quality under training and novel illumination. Dashes denote an evaluation that is not supported by a method.
}
%\vspace{-0.3cm}
\label{tab:quant-comparison}
\begin{tabular}{lrrrrrrrrr}
\toprule
Mode & \multicolumn{7}{c}{Reconstruction} & \multicolumn{2}{c}{Generation} \\
\cmidrule(lr){2-8}
\cmidrule(lr){9-10}
\multirow{2}{*}{\raisebox{-0.7ex}{Evaluation}} & \multicolumn{3}{c}{Medium Properties} &  \multicolumn{3}{c}{\multirow{2}{*}{\raisebox{-2.0ex}{\makecell[c]{Novel-view Synthesis\\ \& Relighting}}}} & \multicolumn{1}{c}{\multirow{2}{*}{\raisebox{-0.7ex}{\makecell[c]{Time}}}} & \multirow{2}{*}{\raisebox{-2.0ex}{\makecell[c]{Train.\\ \,Illum.}}} & \multirow{2}{*}{\raisebox{-2.0ex}{\makecell[c]{Novel\\ \,Illum.}}} \\
\cmidrule(lr){2-4}
& \multicolumn{1}{c}{$\absorptioncoeff$} & \multicolumn{1}{c}{$\scatteringcoeff$} & \multicolumn{1}{c}{$\extinctioncoeff$} \\
\cmidrule(lr){2-2}
\cmidrule(lr){3-3}
\cmidrule(lr){4-4}
\cmidrule(lr){5-7}
\cmidrule(lr){8-8}
\cmidrule(lr){9-9}
\cmidrule(lr){10-10}
Metric & MSE{\tiny$\times 10^2$}$\downarrow$ & MSE{\tiny$\times 10^2$}$\downarrow$ & MSE{\tiny$\times 10^2$}$\downarrow$ & PSNR$\uparrow$ & SSIM$\uparrow$ & LPIPS$\downarrow$ & \multicolumn{1}{c}{hours} & \multicolumn{1}{c}{FID$\downarrow$} & \multicolumn{1}{c}{FID$\downarrow$} \\
\midrule
Diff. Ratio Tracking & 0.87 & 4.06 & 6.88 & 43.3 &  .993 & .035 & 3.0 & --- & --- \\
TensorIR & --- & --- & --- & 11.4 & .594 & .595 & 6.3 & --- & --- \\
\textbf{Ours} & 1.33 & 1.66 & 0.46 & 30.5 & .989 & .054 & 5.5 & 42.3 & 40.8 \\
\bottomrule
\end{tabular}
\end{center}
\end{table*}

\begin{table}
\begin{center} 
\caption
{
    Evaluation of our method for volume reconstruction on scenes with \emph{anisotropic} phase functions using the MSE ($\times 10^2$).
}
%\vspace{-0.3cm}
\label{tab:quant-anisotropic}
\begin{tabular}{lrrrr}
\toprule
& \multicolumn{1}{c}{$\absorptioncoeff$} & \multicolumn{1}{c}{$\scatteringcoeff$} & \multicolumn{1}{c}{$\extinctioncoeff$} & \multicolumn{1}{c}{$\hgparam$} \\
\midrule
\textbf{Ours} & 1.99 & 2.37 & 1.33 & 1.21  \\
\bottomrule
\vspace{-0.7cm}
\end{tabular}
\end{center}
\end{table}

\section{Evaluation}

We first describe our evaluation setup, followed by an evaluation of our approach on volumetric scene reconstruction~(\refSec{eval-recon}) and generation~(\refSec{eval-gen}). We conclude with additional analysis~(\refSec{eval-analysis}).

\noindent
\textbf{Data\quad}
Our work explores a novel formulation for modeling global illumination in volumetric inverse rendering, rather than aiming to develop a robust end-to-end pipeline for imperfect real-world data.
\rev{To our knowledge, real multi-view datasets of volumetric scenes with known illumination do not yet exist.}
Accordingly, we evaluate our approach on synthetic scenes with known camera parameters and illumination conditions, with access to ground-truth volumetric properties for evaluation.
To cover a broad range of volumetric configurations, we construct a custom dataset of 50 scenes. 
Each scene is assembled from a random set of volumetric elements drawn from the Disney Clouds dataset~\cite{10.1145/3130800.3130880}, which spans a continuum of structures from diffuse, low-frequency regions to sharply defined boundaries.
For each selected element, we define RGB absorption and scattering coefficients, \absorptioncoeff and \scatteringcoeff, by scaling its base volumetric density with independently sampled random colors. 
This yields a high dynamic range of \absorptioncoeff and \scatteringcoeff (coefficients take values up to 5, \rev{corresponding to an expected 20 light--volume interactions per path}) and produces diverse combinations of absorption and scattering, resulting in a wide range of optical behaviors spanning weakly absorbing, strongly scattering, and highly opaque elements.
We additionally assign each element a Henyey-Greenstein phase function parameter \hgparam sampled uniformly from $[-0.5, 0.5]$.
A scene is formed by randomly selecting between one and four augmented elements, applying independent random spatial shifts, rotations, and isotropic scalings, and combining them into a single volume.
Illumination is provided by ten environment maps from Polyhaven~\cite{polyhaven}. 
For each scene, we randomly choose two different environment maps and render the scene under both conditions to enable evaluation of relighting. 
Observations are obtained by volumetric path tracing from 500 camera viewpoints placed on a sphere around the volume and oriented toward its center; we reserve 50 views for testing.
In addition to this multi-illumination dataset, we construct a single-illumination variant in which the same environment map is shared across all scenes for training and evaluating the generative setting. 
Finally, to enable direct comparison with baseline implementations that do not support anisotropic scattering, we also provide a variant with isotropic scattering in which all phase functions are fixed to $\hgparam=0$.
All datasets \rev{and the scripts used to generate them} will be released upon publication.

%In addition to this large-scale dataset used for quantitative evaluation, we further consider selected scenes from the \unsure{X dataset}~\cite{fill} that contain near-field light sources \TODO{and other aspects for additional demonstrations}.

\noindent
\textbf{Baselines \quad}
We consider two baseline approaches.
First, we compare against Differential Ratio Tracking (DRT)~\cite{nimier2022unbiased}, which accounts for global illumination by differentiating through stochastic light path sampling, \rev{enabling the optimization of a 3D grid of optical parameters}. 
Like our approach, the method relies on an emission--absorption initialization. 
To ensure a meaningful comparison, we initialize both methods using our emission--absorption solution, which we found yields improved performance for DRT compared to its default setup.
\rev{For quality evaluation, we perform an equal-time comparison by running DRT beyond its recommended iteration count until its total runtime matches that of our method. 
The resulting slightly improved DRT results are those reported. 
For runtime comparisons, however, we use DRT’s recommended hyperparameters.}
DRT's publicly available implementation supports only isotropic phase functions for inverse rendering.

\rev{Our second baseline is an approach based on a learned representation.}
Since we are not aware of a learning-based method that publicly provides a complete implementation for volumetric inverse rendering under global illumination, we compare against TensorIR~\cite{jin2023tensoir} under fixed illumination as a representative baseline.
TensorIR approximates second-order light transport using ray marching through an emissive volume and is not designed to recover physical volumetric optical properties.

%To ensure a fair comparison, we allow all methods to optimize for the same amount of time and verify that each has reached convergence.

%################################################################

\subsection{Reconstruction}
\label{sec:eval-recon}

We evaluate all methods on the large-scale dataset with isotropic scattering using two evaluation protocols.
First, we compute the mean squared error (MSE) of the reconstructed absorption, scattering, and extinction coefficients, as well as the phase function parameters, by densely sampling the corresponding quantities throughout the volume; 
the error of \hgparam is weighted by the local scattering coefficient \scatteringcoeff prior to aggregation.
Second, we evaluate novel-view synthesis on test views under a novel illumination condition.
Images for Differential Ratio Tracking and our approach are synthesized by re-rendering the reconstructed volumes using volumetric path tracing.
We evaluate image quality using PSNR, SSIM~\cite{wang2004image}, and LPIPS~\cite{zhang2018unreasonable} on logarithmically tonemapped images to account for high dynamic range.

\begin{figure*}
    \includegraphics[width=0.99\textwidth]{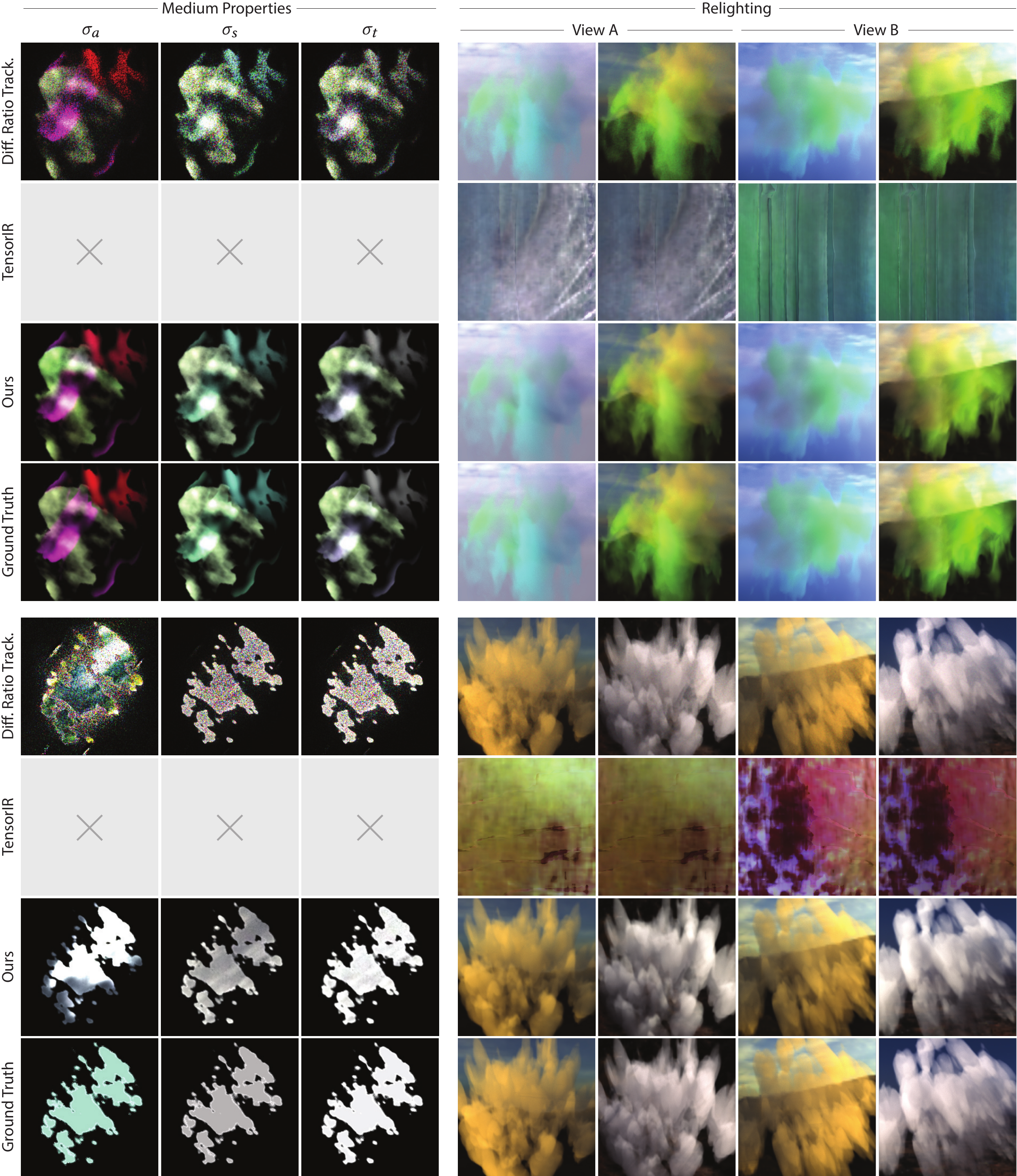}
    \vspace{-0.1cm}
    \caption{
       Reconstruction results on two scenes (row blocks), comparing different methods (rows).
       The first three columns show linearly tonemapped slices through the reconstructed medium properties, namely absorption (\absorptioncoeff), scattering (\scatteringcoeff), and extinction (\extinctioncoeff).
       The remaining columns demonstrate novel-view synthesis under novel illumination.
       TensorIR does not recover volumetric medium properties and is therefore shown only for the image-based comparisons.
    }
    \label{fig:qual-recon}
\end{figure*}

Quantitative results are listed in \refTab{quant-comparison}, left, while qualitative comparisons are shown in \refFig{qual-recon}.
\rev{We observe that our method achieves significantly more accurate recovery of the scattering and extinction coefficients than differentiable stochastic path sampling.}
At the same time, image-level novel-view synthesis quality is highest for the stochastic path-sampling approach, reflecting the benefit of explicit Monte Carlo simulation for optimizing pixel-level appearance.
The learning-based baseline did not produce competitive results on our challenging dataset, despite extensive efforts to ensure a fair evaluation. 
Specifically, we tested configurations using LDR input observations, fixed the environment illumination to the ground truth, and explored a range of training hyperparameters.
As we were able to reproduce the authors' reported results on the datasets used in their evaluation, we hypothesize that the method may not be well suited to the intricate volumetric structures present in our dataset.

%These results highlight the distinction between learning appearance under restricted transport assumptions and learning physical volumes constrained by radiative transfer.

A quantitative evaluation of reconstruction quality for our method with anisotropic scattering is provided in \refTab{quant-anisotropic}. 
We observe that incorporating anisotropic scattering as an additional degree of freedom leads to only a moderate reduction in reconstruction accuracy compared to the isotropic case, while none of the baseline implementations support recovery of anisotropic scattering coefficients.
\refFig{teaser} further demonstrates the accuracy of the recovered scene properties for an anisotropic medium under complex relighting, including strong local light sources.
%\unsure{Please consult our supplemental video for additional results.}
%In\ \refFig{teaser}, we present an additional result for a scene with both environment illumination and near-field light sources.

%################################################################

\subsection{Generation}
\label{sec:eval-gen}

\begin{figure}
    \includegraphics[width=0.99\linewidth]{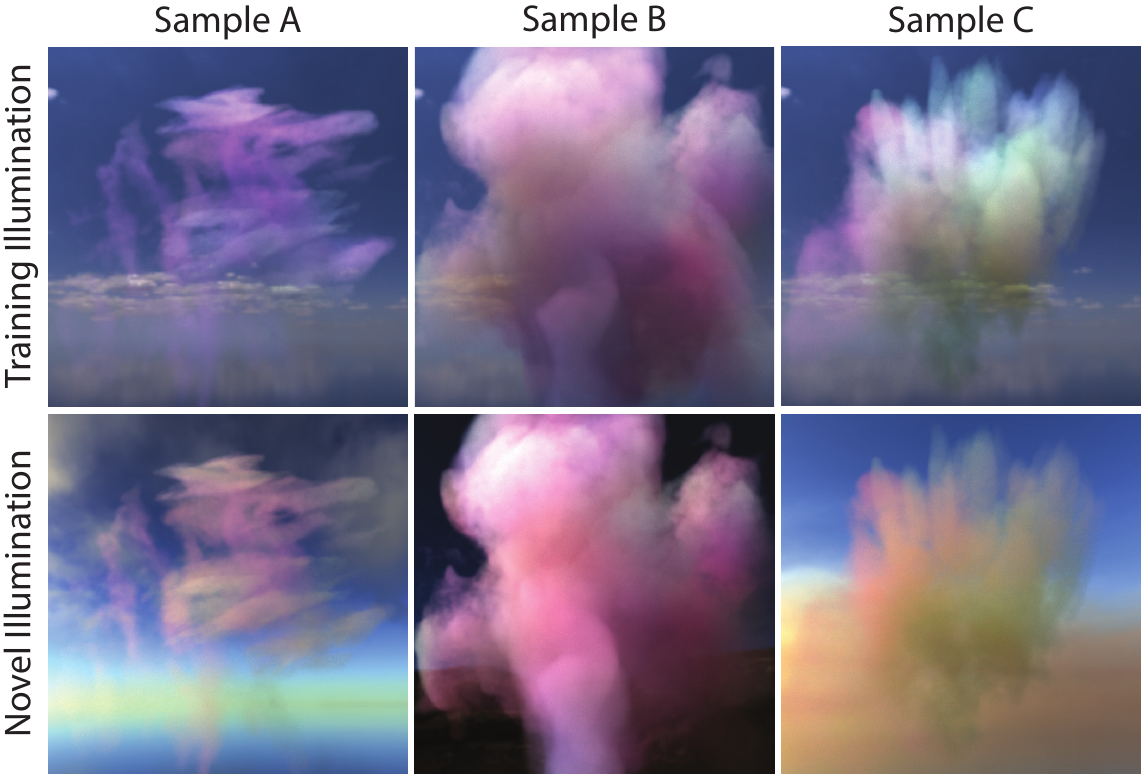}
    \vspace{-0.2cm}
    \caption{
       Three scenes (columns) sampled from our generative model, showing diverse volumetric structures with physically meaningful optical properties, rendered under training (top row) and novel (bottom row) illumination. All images are rendered from the same view.
    }
    \label{fig:qual-generative}
\end{figure}

To evaluate the generative capabilities of our approach, we train a generative model using our large-scale dataset with a single shared illumination condition and a latent dimensionality of $\latentdim = 4$.
We then synthesize 30 scenes by sampling latent codes \latent and render the corresponding scenes under the training illumination.
To assess the faithfulness of the learned distribution over optical properties, we additionally render the same generated scenes under randomly sampled novel illuminations.
In both cases, the distribution of the resulting renderings is compared to that of the corresponding real images -- either from the training set or from the multi-illumination dataset -- using the FID score~\cite{heusel2017gans}.
For both evaluations, we extract random crops from each image to improve the robustness of the distributional comparison.
The corresponding scores are reported in \refTab{quant-comparison}, right, while qualitative samples are shown in \refFig{qual-generative}.
Our results show that the model learns a generative distribution over physical volumetric optical properties -- rather than appearance -- yielding scene samples that support consistent global illumination and relighting, a capability that has not been demonstrated by prior volumetric generative approaches.

%################################################################

\subsection{Ablations}
\label{sec:eval-analysis}

\rev{Our formulation is intentionally minimal, relying on a small set of physically motivated objectives~(\refSec{abl-objectives}) and two generic neural fields for scene representation~(\refSec{abl-rep}), with solutions obtained via optimization~(\refSec{abl-optimization}).}

\subsubsection{Objectives}
\label{sec:abl-objectives}

\begin{figure}
    \includegraphics[width=0.99\linewidth]{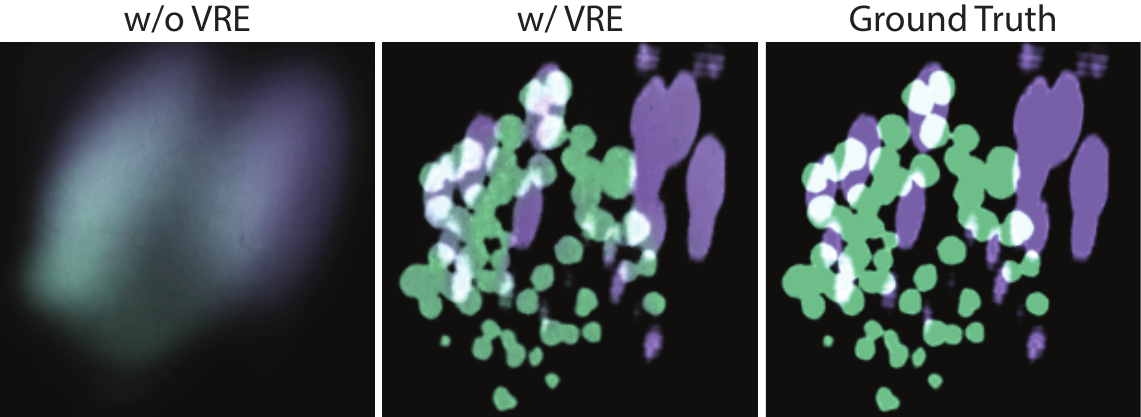}
    \vspace{-0.2cm}
    \caption{
        Reconstruction of scattering parameters \scatteringcoeff (shown here as representative of all recovered volumetric parameters) on a slice through the volume.
        Omitting the VRE objective (\refEq{vre-loss}) leads to overly smooth reconstructions, whereas incorporating it yields results closer to the ground truth.
    }
    \label{fig:ablation}
\end{figure}

Most of our optimization objectives are essential: removing any of the RTE residual, boundary conditions, or observation constraints renders the problem underconstrained and prevents convergence to a meaningful solution. 
\rev{Specifically, this leads to physically inconsistent solutions, neglect of illumination, and disregard of observations, respectively.}
We therefore focus our ablation study on the only component of our formulation that appears optional, namely the VRE-based objective (\refEq{vre-loss}).
As shown in \refFig{ablation}, omitting the VRE term leads to overly smooth reconstructions that lack high-frequency detail.
This behavior is consistent with the known \rev{low-frequency} bias of physics-informed neural networks relying solely on local differential constraints.
\rev{Importantly, this effect is largely orthogonal to architectural choices such as positional encoding~\cite{mildenhall2020nerf,tancik2020fourier}, which only provide the \emph{capacity} to represent high-frequency functions but still require sufficiently direct supervision for this capacity to be utilized.
Incorporating the nonlocal VRE objective effectively provides such supervision and yields substantially sharper and more faithful reconstructions.}

\subsubsection{\rev{Representation}}
\label{sec:abl-rep}

\rev{
We use neural fields to represent both the light field and the volumetric properties.
For reconstruction, the former is inherently a 5D function and thus naturally represented using a neural field, whereas the latter is 3D and therefore, in principle, amenable to straightforward discretization.
However, as shown in \refTab{ablation}, a grid-based representation yields inferior results compared to our neural representation in this setting.
Moreover, grid-based representations are fundamentally incompatible with our generative modeling formulation.
}

\subsubsection{\rev{Optimization}}
\label{sec:abl-optimization}

\rev{Our optimization proceeds in two stages. First, an emission--absorption model is obtained, whose density is subsequently modulated by a function optimized in the second stage to yield the final medium properties.
As shown in \refTab{ablation}, a direct, single-stage optimization yields inferior results compared to our two-stage approach.}

\begin{table}
\begin{center} 
\caption
{
    Quantitative results of ablation studies, measured as MSE~($\times 10^2$) of reconstructed medium properties.
}
%\vspace{-0.3cm}
\label{tab:ablation}
\begin{tabular}{lrrr}
\toprule
& \multicolumn{1}{c}{$\absorptioncoeff$} & \multicolumn{1}{c}{$\scatteringcoeff$} & \multicolumn{1}{c}{$\extinctioncoeff$} \\
\midrule
Grid Representation & 1.98 & 5.20 & 3.43 \\
Single-stage Optimization & 1.68 & 2.94 & 2.60 \\
\textbf{Ours} & 1.33 & 1.66 & 0.46  \\
\bottomrule
\vspace{-0.7cm}
\end{tabular}
\end{center}
\end{table}

%################################################################
%################################################################

\section{Conclusion}

We have demonstrated that, for global-illumination-aware inverse rendering, a global-illumination renderer is not essential.
By representing both the optical properties of participating media and the full light field as neural fields, and by constraining them through predominantly local formulations of radiative transfer, our optimization-based approach enables the reconstruction of high-quality, physically meaningful volumetric parameters.
Beyond inverse rendering, the proposed formulation naturally supports generative modeling of participating media with physical optical properties, learned directly from image observations.

Because both the scene representation and the optimization objective are expressed in terms of principled, physics-based constraints, the framework is agnostic to the dimensionality of the underlying signal and compositional with respect to constraints.
This opens promising directions for extending the approach to spectral and transient light transport, as well as dynamic scenes, while preserving physical consistency.
\revv{Additionally, while our experiments focus on environment illumination, the proposed framework can be extended to other lighting representations, such as local light sources, by incorporating the emission term into the radiative transfer equation. Further, exploring the application of Eq.~\ref{eq:vre-loss} to estimate incident illumination within the volume could lead to hybrid approaches that combine our neural formulation with explicit simulation.}
We believe this perspective provides a flexible foundation for future work on volumetric scene understanding and synthesis.

% Further, exploring the application of Eq.~\ref{eq:vre-loss} to other scenarios, such as incident illumination, could lead to hybrid approaches that combine our neural formulation with explicit simulation.}

%################################################################
%################################################################

% acknowledgments: Felix, Steph

\section*{Acknowledgments}
We thank Felix Mujkanovic and Stephanie Wenxin Liu for their assistance with the evaluation of this work. \\
Open Access funding enabled and organized by Projekt DEAL.

%################################################################
%################################################################

\bibliographystyle{eg-alpha-doi}  
\bibliography{bib}

@article{chu2022physics,
  title={Physics informed neural fields for smoke reconstruction with sparse data},
  author={Chu, Mengyu and Liu, Lingjie and Zheng, Quan and Franz, Aleksandra and Seidel, Hans-Peter and Theobalt, Christian and Zayer, Rhaleb},
  journal={ACM Transactions on Graphics (ToG)},
  volume={41},
  number={4},
  year={2022},
  publisher={ACM New York, NY, USA}
}

@article{yu2024inferring,
  title={Inferring hybrid neural fluid fields from videos},
  author={Yu, Hong-Xing and Zheng, Yang and Gao, Yuan and Deng, Yitong and Zhu, Bo and Wu, Jiajun},
  journal={Advances in Neural Information Processing Systems},
  volume={36},
  year={2024}
}

@article{raissi2019physics,
  title={Physics-informed neural networks: A deep learning framework for solving forward and inverse problems involving nonlinear partial differential equations},
  author={Raissi, Maziar and Perdikaris, Paris and Karniadakis, George E},
  journal={Journal of Computational physics},
  volume={378},
  pages={686--707},
  year={2019},
  publisher={Elsevier}
}

@article{gkioulekas2013inverse,
  title={Inverse volume rendering with material dictionaries},
  author={Gkioulekas, Ioannis and Zhao, Shuang and Bala, Kavita and Zickler, Todd and Levin, Anat},
  journal={ACM Transactions on Graphics (TOG)},
  volume={32},
  number={6},
  pages={1--13},
  year={2013},
  publisher={ACM New York, NY, USA}
}

@article{zheng2021neural,
  title={Neural relightable participating media rendering},
  author={Zheng, Quan and Singh, Gurprit and Seidel, Hans-Peter},
  journal={Advances in Neural Information Processing Systems},
  volume={34},
  year={2021}
}

@inproceedings{srinivasan2021nerv,
  title={Nerv: Neural reflectance and visibility fields for relighting and view synthesis},
  author={Srinivasan, Pratul P and Deng, Boyang and Zhang, Xiuming and Tancik, Matthew and Mildenhall, Ben and Barron, Jonathan T},
  booktitle={Proceedings of the IEEE/CVF conference on computer vision and pattern recognition},
  year={2021}
}

@article{bi2020neural,
  title={Neural reflectance fields for appearance acquisition},
  author={Bi, Sai and Xu, Zexiang and Srinivasan, Pratul and Mildenhall, Ben and Sunkavalli, Kalyan and Ha{\v{s}}an, Milo{\v{s}} and Hold-Geoffroy, Yannick and Kriegman, David and Ramamoorthi, Ravi},
  journal={arXiv preprint arXiv:2008.03824},
  year={2020}
}

@article{zhang2021nerfactor,
  title={Nerfactor: Neural factorization of shape and reflectance under an unknown illumination},
  author={Zhang, Xiuming and Srinivasan, Pratul P and Deng, Boyang and Debevec, Paul and Freeman, William T and Barron, Jonathan T},
  journal={ACM Transactions on Graphics (ToG)},
  volume={40},
  number={6},
  pages={1--18},
  year={2021},
  publisher={ACM New York, NY, USA}
}

@inproceedings{zhang2023nemf,
  title={Nemf: Inverse volume rendering with neural microflake field},
  author={Zhang, Youjia and Xu, Teng and Yu, Junqing and Ye, Yuteng and Jing, Yanqing and Wang, Junle and Yu, Jingyi and Yang, Wei},
  booktitle={Proceedings of the IEEE/CVF International Conference on Computer Vision},
  year={2023}
}

@inproceedings{mildenhall2020nerf,
  title={NeRF: Representing Scenes as Neural Radiance Fields for View Synthesis},
  author={Mildenhall, Ben and Srinivasan, Pratul P and Tancik, Matthew and Barron, Jonathan T and Ramamoorthi, Ravi and Ng, Ren},
  booktitle={European Conference on Computer Vision},
  pages={405--421},
  year={2020},
  organization={Springer}
}

@article{weier2025practical,
  title={Practical Inverse Rendering of Textured and Translucent Appearance},
  author={Weier, Philippe and Riviere, J{\'e}r{\'e}my and Guseinov, Ruslan and Garbin, Stephan and Slusallek, Philipp and Bickel, Bernd and Beeler, Thabo and Vicini, Delio},
  journal={ACM Transactions on Graphics (TOG)},
  volume={44},
  number={4},
  pages={1--16},
  year={2025},
  publisher={ACM New York, NY, USA}
}

@inproceedings{gkioulekas2016evaluation,
  title={An evaluation of computational imaging techniques for heterogeneous inverse scattering},
  author={Gkioulekas, Ioannis and Levin, Anat and Zickler, Todd},
  booktitle={European Conference on Computer Vision},
  pages={685--701},
  year={2016},
  organization={Springer}
}

@article{zhang2019differential,
  title={A differential theory of radiative transfer},
  author={Zhang, Cheng and Wu, Lifan and Zheng, Changxi and Gkioulekas, Ioannis and Ramamoorthi, Ravi and Zhao, Shuang},
  journal={ACM Transactions on Graphics (TOG)},
  volume={38},
  number={6},
  pages={1--16},
  year={2019},
  publisher={ACM New York, NY, USA}
}

@inproceedings{che2020towards,
  title={Towards learning-based inverse subsurface scattering},
  author={Che, Chengqian and Luan, Fujun and Zhao, Shuang and Bala, Kavita and Gkioulekas, Ioannis},
  booktitle={2020 IEEE International Conference on Computational Photography (ICCP)},
  pages={1--12},
  year={2020},
  organization={IEEE}
}

@article{nimier2022unbiased,
  title={Unbiased inverse volume rendering with differential trackers},
  author={Nimier-David, Merlin and M{\"u}ller, Thomas and Keller, Alexander and Jakob, Wenzel},
  journal={ACM Transactions on Graphics (TOG)},
  volume={41},
  number={4},
  pages={1--20},
  year={2022},
  publisher={ACM New York, NY, USA}
}

@article{zhang2021path,
  title={Path-space differentiable rendering of participating media},
  author={Zhang, Cheng and Yu, Zihan and Zhao, Shuang},
  journal={ACM Transactions on Graphics (TOG)},
  volume={40},
  number={4},
  pages={1--15},
  year={2021},
  publisher={ACM New York, NY, USA}
}

@inproceedings{deng2022reconstructing,
  title={Reconstructing translucent objects using differentiable rendering},
  author={Deng, Xi and Luan, Fujun and Walter, Bruce and Bala, Kavita and Marschner, Steve},
  booktitle={ACM SIGGRAPH 2022 Conference Proceedings},
  pages={1--10},
  year={2022}
}

@inproceedings{deng2024reconstructing,
  title={Reconstructing translucent thin objects from photos},
  author={Deng, Xi and Wu, Lifan and Walter, Bruce and Ramamoorthi, Ravi and d'Eon, Eugene and Marschner, Steve and Weidlich, Andrea},
  booktitle={SIGGRAPH Asia 2024 Conference Papers},
  pages={1--11},
  year={2024}
}

@inproceedings{levis2015airborne,
  title={Airborne three-dimensional cloud tomography},
  author={Levis, Aviad and Schechner, Yoav Y and Aides, Amit and Davis, Anthony B},
  booktitle={Proceedings of the IEEE International Conference on Computer Vision},
  pages={3379--3387},
  year={2015}
}

@article{li2018differentiable,
  title={Differentiable monte carlo ray tracing through edge sampling},
  author={Li, Tzu-Mao and Aittala, Miika and Durand, Fr{\'e}do and Lehtinen, Jaakko},
  journal={ACM Transactions on Graphics (TOG)},
  volume={37},
  number={6},
  pages={1--11},
  year={2018},
  publisher={ACM New York, NY, USA}
}

@article{nimier2020radiative,
  title={Radiative backpropagation: An adjoint method for lightning-fast differentiable rendering},
  author={Nimier-David, Merlin and Speierer, S{\'e}bastien and Ruiz, Beno{\^\i}t and Jakob, Wenzel},
  journal={ACM Transactions on Graphics (TOG)},
  volume={39},
  number={4},
  pages={146--1},
  year={2020},
  publisher={ACM New York, NY, USA}
}

@article{zeltner2021monte,
  title={Monte Carlo estimators for differential light transport},
  author={Zeltner, Tizian and Speierer, S{\'e}bastien and Georgiev, Iliyan and Jakob, Wenzel},
  journal={ACM Transactions on Graphics (TOG)},
  volume={40},
  number={4},
  pages={1--16},
  year={2021},
  publisher={ACM New York, NY, USA}
}

@inproceedings{zhang2021physg,
  title={Physg: Inverse rendering with spherical gaussians for physics-based material editing and relighting},
  author={Zhang, Kai and Luan, Fujun and Wang, Qianqian and Bala, Kavita and Snavely, Noah},
  booktitle={Proceedings of the IEEE/CVF conference on computer vision and pattern recognition},
  pages={5453--5462},
  year={2021}
}

@article{boss2021neural,
  title={Neural-pil: Neural pre-integrated lighting for reflectance decomposition},
  author={Boss, Mark and Jampani, Varun and Braun, Raphael and Liu, Ce and Barron, Jonathan and Lensch, Hendrik},
  journal={Advances in Neural Information Processing Systems},
  volume={34},
  pages={10691--10704},
  year={2021}
}

@inproceedings{wu2025pbr,
  title={PBR-NeRF: Inverse Rendering with Physics-Based Neural Fields},
  author={Wu, Sean and Basu, Shamik and Broedermann, Tim and Van Gool, Luc and Sakaridis, Christos},
  booktitle={Proceedings of the Computer Vision and Pattern Recognition Conference},
  pages={10974--10984},
  year={2025}
}

@inproceedings{liang2024gs,
  title={Gs-ir: 3d gaussian splatting for inverse rendering},
  author={Liang, Zhihao and Zhang, Qi and Feng, Ying and Shan, Ying and Jia, Kui},
  booktitle={Proceedings of the IEEE/CVF Conference on Computer Vision and Pattern Recognition},
  pages={21644--21653},
  year={2024}
}

@article{shi2025gir,
  author    = {Shi, Yahao and Wu, Yanmin and Wu, Chenming and Liu, Xing and Zhao, Chen and Feng, Haocheng and Zhang, Jian and Zhou, Bin and Ding, Errui and Wang, Jingdong},
  title     = {GIR: 3D Gaussian Inverse Rendering for Relightable Scene Factorization},
  journal   = {IEEE Transactions on Transactions on Pattern Analysis and Machine Intelligence},
  year      = {2025},
}

@article{dihlmann2024subsurface,
  title={Subsurface Scattering for Gaussian Splatting},
  author={Dihlmann, Jan-Niklas and Majumdar, Arjun and Engelhardt, Andreas and Braun, Raphael and Lensch, Hendrik},
  journal={Advances in Neural Information Processing Systems},
  volume={37},
  pages={121765--121789},
  year={2024}
}

@inproceedings{chen2025gigs,
  title     = {GI-GS: Global Illumination Decomposition on Gaussian Splatting for Inverse Rendering},
  author    = {Chen, Hongze and Lin, Zehong and Zhang, Jun},
  booktitle = {Proceedings of the International Conference on Learning Representations (ICLR)},
  year      = {2025}
}

@inproceedings{jin2023tensoir,
  title={Tensoir: Tensorial inverse rendering},
  author={Jin, Haian and Liu, Isabella and Xu, Peijia and Zhang, Xiaoshuai and Han, Songfang and Bi, Sai and Zhou, Xiaowei and Xu, Zexiang and Su, Hao},
  booktitle={Proceedings of the IEEE/CVF Conference on Computer Vision and Pattern Recognition},
  pages={165--174},
  year={2023}
}

@book{pharr2023physically,
  title={Physically based rendering: From theory to implementation},
  author={Pharr, Matt and Jakob, Wenzel and Humphreys, Greg},
  year={2023},
  publisher={MIT Press}
}

@article{vicini2021path,
  title={Path replay backpropagation: Differentiating light paths using constant memory and linear time},
  author={Vicini, Delio and Speierer, S{\'e}bastien and Jakob, Wenzel},
  journal={ACM Transactions on Graphics (TOG)},
  volume={40},
  number={4},
  pages={1--14},
  year={2021},
  publisher={ACM New York, NY, USA}
}

@article{nicolet2023recursive,
  title={Recursive control variates for inverse rendering},
  author={Nicolet, Baptiste and Rousselle, Fabrice and Novak, Jan and Keller, Alexander and Jakob, Wenzel and M{\"u}ller, Thomas},
  journal={ACM Transactions on Graphics (TOG)},
  volume={42},
  number={4},
  pages={1--13},
  year={2023},
  publisher={ACM New York, NY, USA}
}

@article{nimier2019mitsuba,
  title={Mitsuba 2: A retargetable forward and inverse renderer},
  author={Nimier-David, Merlin and Vicini, Delio and Zeltner, Tizian and Jakob, Wenzel},
  journal={ACM Transactions on Graphics (ToG)},
  volume={38},
  number={6},
  pages={1--17},
  year={2019},
  publisher={ACM New York, NY, USA}
}

@inproceedings{xie2022neural,
  title={Neural fields in visual computing and beyond},
  author={Xie, Yiheng and Takikawa, Towaki and Saito, Shunsuke and Litany, Or and Yan, Shiqin and Khan, Numair and Tombari, Federico and Tompkin, James and Sitzmann, Vincent and Sridhar, Srinath},
  booktitle={Computer graphics forum},
  volume={41},
  number={2},
  pages={641--676},
  year={2022},
  organization={Wiley Online Library}
}

@article{hadadan2021neural,
  title={Neural radiosity},
  author={Hadadan, Saeed and Chen, Shuhong and Zwicker, Matthias},
  journal={ACM Transactions on Graphics (TOG)},
  volume={40},
  number={6},
  pages={1--11},
  year={2021},
  publisher={ACM New York, NY, USA}
}

@inproceedings{hadadan2023inverse,
  title={Inverse global illumination using a neural radiometric prior},
  author={Hadadan, Saeed and Lin, Geng and Nov{\'a}k, Jan and Rousselle, Fabrice and Zwicker, Matthias},
  booktitle={ACM SIGGRAPH 2023 Conference Proceedings},
  pages={1--11},
  year={2023}
}

@article{li2024tensosdf,
  title={Tensosdf: Roughness-aware tensorial representation for robust geometry and material reconstruction},
  author={Li, Jia and Wang, Lu and Zhang, Lei and Wang, Beibei},
  journal={ACM Transactions on Graphics (TOG)},
  volume={43},
  number={4},
  pages={1--13},
  year={2024},
  publisher={ACM New York, NY, USA}
}

@inproceedings{yao2022neilf,
  title={Neilf: Neural incident light field for physically-based material estimation},
  author={Yao, Yao and Zhang, Jingyang and Liu, Jingbo and Qu, Yihang and Fang, Tian and McKinnon, David and Tsin, Yanghai and Quan, Long},
  booktitle={European conference on computer vision},
  pages={700--716},
  year={2022},
  organization={Springer}
}

@inproceedings{Worchel2025RadiativeBackpropagation,
  author    = {Worchel, Markus and Finnendahl, Ugo and Alexa, Marc},
  title     = {Radiative Backpropagation with Non-Static Geometry},
  booktitle = {Proc. EGSR},
  year      = {2025},
}

@article{mishra2021physics,
  title={Physics informed neural networks for simulating radiative transfer},
  author={Mishra, Siddhartha and Molinaro, Roberto},
  journal={Journal of Quantitative Spectroscopy and Radiative Transfer},
  volume={270},
  pages={107705},
  year={2021},
  publisher={Elsevier}
}

@Article{violante2024lighting_3D_cars,
      author       = {Violante, Nicolás and Gauthier, Alban and Diolatzis, Stavros and Leimkühler, Thomas and Drettakis, George},
      title        = {Physically-Based Lighting for 3D Generative Models of Cars},
      journal      = {Computer Graphics Forum (Proceedings of the Eurographics Conference)},
      number       = {2},
      volume       = {43},
      month        = {April},
      year         = {2024}
}

@inproceedings{lyu2022nrtf,
title = {Neural Radiance Transfer Fields for Relightable Novel-view Synthesis with Global Illumination},
author = {Lyu, Linjie and Tewari, Ayush and Leimk{\"u}hler, Thomas and Habermann, Marc and Theobalt, Christian},
year = {2022},
booktitle={ECCV},
}

@article{pan2021shading,
  title={A shading-guided generative implicit model for shape-accurate 3d-aware image synthesis},
  author={Pan, Xingang and Xu, Xudong and Loy, Chen Change and Theobalt, Christian and Dai, Bo},
  journal={Advances in Neural Information Processing Systems},
  volume={34},
  pages={20002--20013},
  year={2021}
}

@inproceedings{diolatzis2023mesogan,
  title={Mesogan: Generative neural reflectance shells},
  author={Diolatzis, Stavros and Novak, Jan and Rousselle, Fabrice and Granskog, Jonathan and Aittala, Miika and Ramamoorthi, Ravi and Drettakis, George},
  booktitle={Computer Graphics Forum},
  volume={42},
  number={6},
  pages={e14846},
  year={2023},
  organization={Wiley Online Library}
}

@article{jiang2023nerffacelighting,
  title={Nerffacelighting: Implicit and disentangled face lighting representation leveraging generative prior in neural radiance fields},
  author={Jiang, Kaiwen and Chen, Shu-Yu and Fu, Hongbo and Gao, Lin},
  journal={ACM Transactions on Graphics},
  volume={42},
  number={3},
  pages={1--18},
  year={2023},
  publisher={ACM New York, NY, USA}
}

@article{zucker2025physics,
  title={Physics-informed neural networks for modeling atmospheric radiative transfer},
  author={Zucker, Shai and Batenkov, Dmitry and Rozenhaimer, Michal Segal},
  journal={Journal of Quantitative Spectroscopy and Radiative Transfer},
  volume={331},
  pages={109253},
  year={2025},
  publisher={Elsevier}
}

@article{riganti2023auxiliary,
  title={Auxiliary physics-informed neural networks for forward, inverse, and coupled radiative transfer problems},
  author={Riganti, Roberto and Negro, L Dal},
  journal={Applied Physics Letters},
  volume={123},
  number={17},
  year={2023},
  publisher={AIP Publishing}
}

@InProceedings{chen2023fantasia3d,
  author={Chen, Rui and Chen, Yongwei and Jiao, Ningxin and Jia, Kui},
  title={Fantasia3D: Disentangling Geometry and Appearance for High-quality Text-to-3D Content Creation},
  booktitle={Proceedings of the IEEE/CVF International Conference on Computer Vision (ICCV)},
  month={October},
  year={2023},
  pages={22246-22256}
}

@article{zhang2024dreammat,
  title={Dreammat: High-quality pbr material generation with geometry-and light-aware diffusion models},
  author={Zhang, Yuqing and Liu, Yuan and Xie, Zhiyu and Yang, Lei and Liu, Zhongyuan and Yang, Mengzhou and Zhang, Runze and Kou, Qilong and Lin, Cheng and Wang, Wenping and others},
  journal={ACM Transactions on Graphics (TOG)},
  volume={43},
  number={4},
  pages={1--18},
  year={2024},
  publisher={ACM New York, NY, USA}
}

@inproceedings{henzler2019escaping,
  title={Escaping plato's cave: 3d shape from adversarial rendering},
  author={Henzler, Philipp and Mitra, Niloy J and Ritschel, Tobias},
  booktitle={Proceedings of the IEEE/CVF International Conference on Computer Vision},
  pages={9984--9993},
  year={2019}
}

@inproceedings{chan2022efficient,
  title={Efficient geometry-aware 3d generative adversarial networks},
  author={Chan, Eric R and Lin, Connor Z and Chan, Matthew A and Nagano, Koki and Pan, Boxiao and De Mello, Shalini and Gallo, Orazio and Guibas, Leonidas J and Tremblay, Jonathan and Khamis, Sameh and others},
  booktitle={Proceedings of the IEEE/CVF conference on computer vision and pattern recognition},
  pages={16123--16133},
  year={2022}
}

@inproceedings{muller2023diffrf,
  title={Diffrf: Rendering-guided 3d radiance field diffusion},
  author={M{\"u}ller, Norman and Siddiqui, Yawar and Porzi, Lorenzo and Bulo, Samuel Rota and Kontschieder, Peter and Nie{\ss}ner, Matthias},
  booktitle={Proceedings of the IEEE/CVF Conference on Computer Vision and Pattern Recognition},
  year={2023}
}

@inproceedings{chen2023single,
  title={Single-stage diffusion nerf: A unified approach to 3d generation and reconstruction},
  author={Chen, Hansheng and Gu, Jiatao and Chen, Anpei and Tian, Wei and Tu, Zhuowen and Liu, Lingjie and Su, Hao},
  booktitle={Proceedings of the IEEE/CVF international conference on computer vision},
  pages={2416--2425},
  year={2023}
}

@Article{kerbl3Dgaussians,
      author       = {Kerbl, Bernhard and Kopanas, Georgios and Leimk{\"u}hler, Thomas and Drettakis, George},
      title        = {3D Gaussian Splatting for Real-Time Radiance Field Rendering},
      journal      = {ACM Transactions on Graphics},
      number       = {4},
      volume       = {42},
      month        = {July},
      year         = {2023}
}

@article{kajiya1984ray,
  title={Ray tracing volume densities},
  author={Kajiya, James T and Von Herzen, Brian P},
  journal={ACM SIGGRAPH computer graphics},
  volume={18},
  number={3},
  pages={165--174},
  year={1984},
  publisher={ACM New York, NY, USA}
}

@article{rng2025,
  author={Jiahui Fan and Fujun Luan and Jian Yang and Milos Hasan and Beibei Wang},
  title={RNG: Relightable Neural Gaussians},
  year={2025},
  journal={Proceedings of CVPR 2025},
}

@book{chandrasekhar1960radiative,
  title={Radiative Transfer},
  author={Chandrasekhar, Subrahmanyan},
  year={1960},
  publisher={Courier Corporation}
}

@article{henyey1941diffuse,
  title={Diffuse radiation in the galaxy},
  author={Henyey, Louis G and Greenstein, Jesse Leonard},
  journal={Astrophysical Journal, vol. 93, p. 70-83 (1941).},
  volume={93},
  pages={70--83},
  year={1941}
}

@article{wang2022when,
  title={When and why PINNs fail to train: A neural tangent kernel perspective},
  author={Wang, Sifan and Yu, Xinling and Perdikaris, Paris},
  journal={Journal of Computational Physics},
  volume={449},
  pages={110768},
  year={2022},
  publisher={Elsevier}
}

@inproceedings{rahaman2019spectral,
  title={On the spectral bias of neural networks},
  author={Rahaman, Nasim and Baratin, Aristide and Arpit, Devansh and Draxler, Felix and Lin, Min and Hamprecht, Fred and Bengio, Yoshua and Courville, Aaron},
  booktitle={International conference on machine learning},
  pages={5301--5310},
  year={2019},
  organization={PMLR}
}

@inproceedings{kingma2014adam,
  author={Diederik P. Kingma and Jimmy Ba},
  booktitle={International Conference on Learning Representations (ICLR)},
  title={Adam: {A} Method for Stochastic Optimization},
  year={2015},
}

@inproceedings{park2019deepsdf,
  title={Deepsdf: Learning continuous signed distance functions for shape representation},
  author={Park, Jeong Joon and Florence, Peter and Straub, Julian and Newcombe, Richard and Lovegrove, Steven},
  booktitle={Proceedings of the IEEE/CVF conference on computer vision and pattern recognition},
  pages={165--174},
  year={2019}
}

@article{sitzmann2021light,
  title={Light field networks: Neural scene representations with single-evaluation rendering},
  author={Sitzmann, Vincent and Rezchikov, Semon and Freeman, Bill and Tenenbaum, Josh and Durand, Fredo},
  journal={Advances in Neural Information Processing Systems},
  volume={34},
  year={2021}
}

@article{wang2004image,
  title={Image quality assessment: from error visibility to structural similarity},
  author={Wang, Zhou and Bovik, Alan C and Sheikh, Hamid R and Simoncelli, Eero P},
  journal={IEEE transactions on image processing},
  volume={13},
  number={4},
  pages={600--612},
  year={2004},
  publisher={IEEE}
}

@inproceedings{zhang2018unreasonable,
  title={The unreasonable effectiveness of deep features as a perceptual metric},
  author={Zhang, Richard and Isola, Phillip and Efros, Alexei A and Shechtman, Eli and Wang, Oliver},
  booktitle={Proceedings of the IEEE conference on computer vision and pattern recognition},
  pages={586--595},
  year={2018}
}

@article{heusel2017gans,
  title={Gans trained by a two time-scale update rule converge to a local nash equilibrium},
  author={Heusel, Martin and Ramsauer, Hubert and Unterthiner, Thomas and Nessler, Bernhard and Hochreiter, Sepp},
  journal={Advances in neural information processing systems},
  volume={30},
  year={2017}
}

@article{paszke2017automatic,
  title={Automatic differentiation in pytorch},
  author={Paszke, Adam and Gross, Sam and Chintala, Soumith and Chanan, Gregory and Yang, Edward and DeVito, Zachary and Lin, Zeming and Desmaison, Alban and Antiga, Luca and Lerer, Adam},
  year={2017}
}

@article{muller2022instant,
  title={Instant neural graphics primitives with a multiresolution hash encoding},
  author={M{\"u}ller, Thomas and Evans, Alex and Schied, Christoph and Keller, Alexander},
  journal={ACM transactions on graphics (TOG)},
  volume={41},
  number={4},
  pages={1--15},
  year={2022},
  publisher={ACM New York, NY, USA}
}

@article{sitzmann2020implicit,
  title={Implicit neural representations with periodic activation functions},
  author={Sitzmann, Vincent and Martel, Julien and Bergman, Alexander and Lindell, David and Wetzstein, Gordon},
  journal={Advances in neural information processing systems},
  volume={33},
  pages={7462--7473},
  year={2020}
}

@inproceedings{azinovic2019inverse,
  title={Inverse path tracing for joint material and lighting estimation},
  author={Azinovic, Dejan and Li, Tzu-Mao and Kaplanyan, Anton and Nie{\ss}ner, Matthias},
  booktitle={Proceedings of the IEEE/CVF conference on computer vision and pattern recognition},
  pages={2447--2456},
  year={2019}
}

@article{10.1145/3130800.3130880,
author = {Kallweit, Simon and M\"{u}ller, Thomas and Mcwilliams, Brian and Gross, Markus and Nov\'{a}k, Jan},
title = {Deep scattering: rendering atmospheric clouds with radiance-predicting neural networks},
year = {2017},
issue_date = {December 2017},
publisher = {Association for Computing Machinery},
address = {New York, NY, USA},
volume = {36},
number = {6},
issn = {0730-0301},
journal = {ACM Trans. Graph.},
month = nov,
articleno = {231},
numpages = {11},
}

@online{polyhaven,
  title   = {Poly\,Haven: The Public 3D Asset Library},
  url     = {https://polyhaven.com/},
  urldate = {2026-01-19},
  author  = {{Poly\,Haven}},
  year={2026}
}

@article{tancik2020fourier,
  title={Fourier features let networks learn high frequency functions in low dimensional domains},
  author={Tancik, Matthew and Srinivasan, Pratul and Mildenhall, Ben and Fridovich-Keil, Sara and Raghavan, Nithin and Singhal, Utkarsh and Ramamoorthi, Ravi and Barron, Jonathan and Ng, Ren},
  journal={Advances in neural information processing systems},
  volume={33},
  pages={7537--7547},
  year={2020}
}

@inproceedings{pauly2000metropolis,
  title={Metropolis light transport for participating media},
  author={Pauly, Mark and Kollig, Thomas and Keller, Alexander},
  booktitle={Eurographics Workshop on Rendering Techniques},
  pages={11--22},
  year={2000},
  organization={Springer}
}

@article{keller2019integral,
  title={Integral equations and machine learning},
  author={Keller, Alexander and Dahm, Ken},
  journal={Mathematics and Computers in Simulation},
  volume={161},
  pages={2--12},
  year={2019},
  publisher={Elsevier}
}

@article{mccormick1992inverse,
  title={Inverse radiative transfer problems: a review},
  author={McCormick, NJ},
  journal={Nuclear science and Engineering},
  volume={112},
  number={3},
  pages={185--198},
  year={1992},
  publisher={Taylor \& Francis}
}

@article{adams2002fast,
  title={Fast iterative methods for discrete-ordinates particle transport calculations},
  author={Adams, Marvin L and Larsen, Edward W},
  journal={Progress in nuclear energy},
  volume={40},
  number={1},
  pages={3--159},
  year={2002},
  publisher={Elsevier}
}

@article{bangerth2008adaptive,
  title={Adaptive finite element methods for the solution of inverse problems in optical tomography},
  author={Bangerth, Wolfgang and Joshi, Amit},
  journal={Inverse Problems},
  volume={24},
  number={3},
  pages={034011},
  year={2008}
}

@article{abdoulaev2005optical,
  title={Optical tomography as a PDE-constrained optimization problem},
  author={Abdoulaev, Gassan S and Ren, Kui and Hielscher, Andreas H},
  journal={Inverse Problems},
  volume={21},
  number={5},
  pages={1507--1530},
  year={2005}
}

@article{stamnes2000disort,
  title={DISORT, a general-purpose Fortran program for discrete-ordinate-method radiative transfer in scattering and emitting layered media: documentation of methodology},
  author={Stamnes, Knut and Tsay, Si-Chee and Wiscombe, Warren and Laszlo, Istvan},
  year={2000},
  publisher={Tech. rep., Dept. of Physics and Engineering Physics, Stevens Institute of~…}
}

@inproceedings{levis2017multiple,
  title={Multiple-scattering microphysics tomography},
  author={Levis, Aviad and Schechner, Yoav Y and Davis, Anthony B},
  booktitle={Proceedings of the IEEE Conference on Computer Vision and Pattern Recognition},
  pages={6740--6749},
  year={2017}
}
 
\end{document}